\documentclass[]{JHEP}
\usepackage{amssymb}
\usepackage{amsfonts}
\usepackage{latexsym}
\usepackage{amsmath}
\usepackage{epsfig}
\def\beq{\begin{equation}}
\def\bea{\begin{eqnarray}}
\def\eea{\end{eqnarray}}
\newcommand{\ena}{\end{eqnarray}}
\newcommand{\no}{\noindent}
\newcommand{\nb}{\nonumber}

\newcommand{\de}{\partial}
\newcommand{\ha}{\frac{1}{2}}
\def\la{\phi_*}
\def\s{\star}
\def\z{\bar{z}}
\def\addot{\ddot{a}}
\def\mdot{\dot{m}_{3/2}}

\def\t{\Theta}

\title{The Non-commutative Brane World}
\author{Luigi Pilo and Antonio Riotto\\ 
Scuola Normale Superiore, Pisa, Italy\\ 
INFN, Sezione di Pisa, Pisa Italy\\ E--mail:
\email{pilo@cibs.sns.it}, \email{riotto@cibs.sns.it}}

\abstract{
We propose a new higher-dimensional mechanism 
to localize scalar fields as well as fermionic and gauge fields. The underlying
theory is 
a six-dimensional non-commutative field theory where  non-commutativity
is allowed along  two extra infinite spatial dimensions and the 
four-dimensional brane is provided by 
a scalar  soliton living in the non-commutative space.
Making use of the powerful  
 correspondence between non-commutative 
coordinates and operators on a single particle Hilbert space, we show that   
the non-commutative brane world 
admits localized chiral fermions and it ensures
the localization of massless gauge fields. It may also   give rise to
a variety of 
different  low-energy spectra since  the localized zero mode
may come along  either with   a discrete tower of  
degenerate heavy states or  with a tower of
Kaluza-Klein heavy states, or it may even be  the only state in the low-energy
spectrum.}

\keywords{Non-commutative field theory, extra-dimensions}
\preprint{SNS-PH/00-18}

\begin{document}
\section{Introduction}
The capability  of confining Standard Model fields on
$(3+1)$-dimensional subspaces (branes) \cite{brane}
of a higher-dimensional manifold has recently given rise to
    scenarios where the size of the extra dimensions may be 
 much larger than the tiny four-dimensional Planck length.
In models where the space-time geometry
is of a simple factorizable form, the space of extra dimensions -- the 
bulk -- may be compact and perhaps as large as a millimeter \cite{add}.
If the space-time geometry has a non-factorizable form, the extra
dimensions may be warped and non-compact \cite{rs}.  
In particular, the possibility of warped non-compact extra
dimensions has extended our intuition about how extra spatial
dimensions are manifest in four-dimensional effective field theories
by showing that even if gravity propagates in non-compact higher
dimensional spaces, four-dimensional observers may still empirically
deduce a four-dimensional Newton's law.
 
In this paper we will show that the four-dimensional 
localization of scalar, fermionic and
gauge fields may occur in 
 a  six-dimensional non-commutative field theory  where the 
non-commutativity is present in  two extra non-compact spatial dimensions.
Indeed, we find that the magic of  
non-commutativity facilitates by a great deal the  study of localization
of fields and may give rise to many and different kinds of spectra in the 
low-energy four-dimensional theory.

Non-commutative field theories have  turned out to be
very relevant for a broad  variety of unexpected applications.
They are  nonlocal field theories where locality breaks down 
at short
distances and they might provide new  insights 
into the issue of  non-locality in
quantum gravity. The study of the perturbative aspects of
 non-commutative field theories has revealed 
an intriguing mixing of the UV and IR arising from  
non-planar divergent diagrams which are  interpreted  as IR divergences 
\cite{con}.
It has also been understood that
non-commutative gauge theories arise from a limit of string theory
\cite{cds,dh,schom,sw}.

More recently, soliton solutions of scalar non-commutative  field theories
have been constructed \cite{stro}. These
solutions are solitons whose size 
is set by the scale of non-commutativity and 
play an important
role in constructing D-branes
as non-commutative solitons of the tachyon field of open string
theory \cite{dmr,harvey,wit}. 

We start from a   non-commutative field theory of a single real 
scalar $\Phi$
in a six  dimensional space where 
non-commutativity is present along 
 the two extra infinite spatial dimensions, {\it i.e.}
$M_4 \times \mathbb{R}_\s^2$, where $M_4$ denotes the four-dimensional
Minkowski space-time. 
The four-dimensional brane is given by 
             a  radially symmetric soliton 
 in  $\mathbb{R}_\s^2$ and we study
the localization of fields on such a brane taking the limit 
in which the non-commutative length is much larger than the inverse of the
fundamental mass scale in 6D.  The 6D action is initially invariant
under a global $U(\infty)$ symmetry which is broken down to $U(\infty-1)\otimes
U(1)$ by the soliton.

Our construction of the non-commutative brane world 
exploits the powerful connection
between non-commutative coordinates and operators in single particle 
quantum mechanics. Under this correspondence, 
the $\star$-product (the generalization in non-commutative spaces 
of the usual product)
maps onto usual operator multiplication and 
the equation of motions translates into algebraic operator equations.
The four-dimensional  action and the corresponding 
low-energy spectrum can be easily computed by carrying out   a the trace over 
operators and computations are facilitated by the fact that
the non-commutative soliton $\Phi_0$ acts like a
 projector operator in the Hilbert
space of the single particle 
quantum mechanics.
Fields interacting with the soliton may be decomposed
from the four-dimensional point of view along a
a complete and discrete set of orthogonal functions in $\mathbb{R}_\s^2$
and the presence of 
the soliton projects out
most of the unwanted modes,  leaving behind a massless  mode.

A remarkable feature of the resulting 
 four-dimensional action describing the dynamics of  the localized zero
modes 
 is its insensitivity to the details of the underlying non-commutative 
 field theory giving rise
to it.   The localization of a generic field interacting with the
non-commutative soliton   is   due entirely 
to its six-dimensional  kinetic term and to the
fact that the soliton acts like a four-dimensional warp-factor.
The 4D action inherits informations about the non-commutative
soliton only in the point where the soliton field is maximized. This sounds
quite intriguing if we think that the soliton is much broader than the
inverse of the fundamental mass scale of the theory.

Depending upon the kind of interaction between 
a given six-dimensional field and the scalar soliton, different
four-dimensional spectra may arise. In particular,  
the localized zero mode
of a scalar field may come  with either  a discrete tower of  
degenerate heavy states or  with a tower of
Kaluza-Klein heavy states, or it may be the only state in the low-energy
spectrum! We will also show that the non-commutative brane world 
admits localized chiral fermions and ensures
the localization of massless gauge fields once 
   the initial global  $U(\infty)$ symmetry is promoted 
to a non-commutative $U(1)$ gauge symmetry.

The paper is organized as follows. In section 2 we first
review a few  
crucial properties of non-commutative field theories and their solitons.
In section 3 we present our results for the localization of scalar fields,
while localization of fermions is studied in section 4. In section 5
we  see how gauge massless fields descend from  
the soliton world-volume. Finally, in section 6 we present our conclusions
and comment about possible  directions for future work.  

\section{Some Generalities}

\subsection{The Non-commutative Soliton}

As we announced in the introduction, 
our starting point is a  non-commutative field theory of a single real 
scalar $\Phi$
in a six  dimensional space with
non-commutativity in the two extra infinite spatial dimensions, {\it i.e.}
$M_4 \times \mathbb{R}_\s^2$.
 
We will denote the complete set of space-time coordinates
by $x^\mu$ ($\mu=0,\cdots,5$), the non-commutative spatial directions by $x^i$ 
($i=4,5$) and the 
remaining directions by $x^a$ ($a=0,\cdots,3$).

The spatial coordinates $x^4$ and $x^5$  
commute according to 
\beq
\left[x^4,x^5\right] \, = \, i \, \Theta
\end{equation}
and can be  parametrized in terms of the 
complex coordinates $z=(x^4+ix^5)/\sqrt{2}$ and $\bar{z}=(x^4-ix^5)/\sqrt{2}$.

\no
Fields in the non-local action are intended to be 
multiplied using the Moyal star
product,
\beq
\left(f \star g \right)(z, \bar{z}) \, = \, e^{ {\t \over 2}
\left( \partial_{z}\partial_{\bar{z}^{\prime}}
  -\partial_{z^{\prime}}\partial_{\bar{z}} \right)}
f(z, \bar{z})g(z^{\prime}, \bar{z}^{\prime})
\left. \right|_{z_=z^{\prime}} \quad . 
\label{star}
\end{equation}

There is a useful one-to-one correspondence between functions 
$f(z,\bar{z})$
on the non-commutative $\mathbb{R}_\s^2$, thought of as 
the phase space of a particle in one dimension, and operators 
$O_f(\hat{z},\hat{\bar{z}})$ acting on the Hilbert space ${\cal H}$
of a quantum system with one degree of freedom.  Multiplication by 
the $\star$-product goes over to operator multiplication, 
and integration over  $\mathbb{R}_\s^2$ corresponds to taking 
the trace  over the Hilbert space, 
\beq
\label{cor}
f \star g \leftrightarrow \hat{A} \hat{B}~, ~~~~~~ 
{1 \over 2\pi \Theta} \int \! d^2z \, f(z,\bar{z}) \leftrightarrow {\rm Tr}
~O_f(\hat{z},\hat{\bar{z}}) \quad .
\end{equation} 
With this prescription the Moyal product of functions is isomorphic
to ordinary operator multiplication
\beq
O_f \cdot O_g\leftrightarrow O_{f\s g} \quad .
\end{equation}
Consider now the action of  a real scalar field $\Phi$ in $\mathbb{R}^6_\s$
\beq
\label{action}
S \, = \, \int d^6x \left[\ha \, \partial_\mu \Phi \, \partial^\mu \Phi \, - \,  
V(\Phi)\right] \quad .
\end{equation}
The correspondence (\ref{cor}) was exploited in \cite{stro} to construct
general soliton solutions in $\mathbb{R}_\s^2$ in the limit of large 
non-commutativity.
If we  render  the coordinates $z$ and $\z$ dimensionless through
the scaling 
\beq
z \, = \, w \, \sqrt{\Theta}, \qquad  
\bar{z} \, = \,  \bar{w} \, \sqrt{\Theta} \; ,
\end{equation}
the  $\star$-product (\ref{star})
will have no $\Theta$, {\it i.e.} it will be given by (\ref{star})
  with $\Theta=1$.
Written in rescaled coordinates, the dependence
on $\Theta$ in the two-dimensional part of the 
action is entirely in front of the potential term
\beq
\int d^2w \left[- \ha \, \partial_w \Phi \, \partial_{\bar{w}} \Phi \, - \, \Theta \, V(\Phi)\right] \quad .
\end{equation}
In the limit $\Theta\rightarrow \infty$\footnote{This limit
should be intended as  $\Theta M_f^2\rightarrow \infty$ where $M_f$ is the
fundamental scale of the 6D-theory.}
with $V(\Phi)$ held fixed, the kinetic
term is negligible compared to $V(\Phi)$, at least for field
configurations varying over sizes of order one in the new coordinates.

The equation of motion for static solitons is therefore 
\beq
\label{ext}
\frac{dV(\Phi)}{d\Phi} \, = \, 0 \quad . 
\end{equation}
Localized soliton solutions to this equation exist due to the presence of  
the $\star$-product \cite{stro}. The construction  relies on the existence of  
functions $\Phi$ satisfying
\beq
\label{con}
\Phi_0 \star \Phi_0 \, = \, \Phi_0. 
\end{equation}
In such a case, the following crucial property holds
\beq
\label{crucial}
f(\la \Phi_0) \, = \, f(\la) \, \Phi_0 \; , 
\end{equation}
for any function $f$ of the form 
$f(\Phi_0)= \sum_{n=1}^\infty a_n \Phi_0^n$.   
In particular, 
\beq
\left. \frac{dV(\Phi)}{d\Phi}\right|_{\Phi=\la\Phi_0} \, = \, 
\left. \left(\frac{dV(\Phi)}{d\Phi}\right|_{\Phi=\la}\right)
\Phi_0
\end{equation}
and Eq. (\ref{ext}) is solved by choosing $\la$ to be an extremum of 
$V(\Phi)$. 

\no
The simplest function satisfying (\ref{con}) is the Gaussian (in the 
non-rescaled coordinates)
\beq
\Phi_0(r) \, = \, 2 \, e^{-r^2/\Theta},~~~  r^2 \, = \, x_4^2 +x_5^2
\end{equation}
and the soliton solution is
\beq
\label{sol}
\Phi \, = \, \la~\Phi_0(r) \; ,
\end{equation}
where $\la$ is an extremum of $V(\Phi)$,   
$V^{\prime}(\la)=0$.
The non-locality of the $\star$-product makes it possible
for a lump of approximately size $\sqrt{\Theta}$ to square itself
under the $\star$-product  and the resulting object is a $3+1$ dimensional
soliton that we will identify with the brane world where matter fields live.
Note that the soliton solution asymptotically approaches the 
value $\Phi=0$.

The soliton solution is stable \cite{stro}. Indeed, suppose 
there are two solutions
to $V^\prime(\la)=0$, one for $(\phi_*)_1=0$ and $V[(\la)_1]=0$ and one
for $(\phi_*)_1$ and $V[(\la)_2]\neq 0$ and 
suppose they are separated by a finite
barrier.
Far away from the origin,
$\Phi_0(r)=0$, but near $r=0$,  $\Phi_0(r)$ is in the vicinity of the
second vacuum, $\la=(\la)_2$.
The solution corresponds to a bubble of false vacuum. The area of the
bubble is of order one (or $\Theta$ in the non-rescaled coordinates).
In a commutative theory such a bubble would decay by shrinking to zero
size. Non-commutativity prevents the bubble from shrinking to
a spatial size smaller than
$\sqrt{\Theta}$. In order to decay, $\Phi_0$ would have to 
scale to zero, a  process which is
classically forbidden \cite{stro}. Of course, the soliton is unstable
against quantum decay, but the decay rate can be made arbitrarily small
by adjusting the parameters of the potential.

\subsection{The Symmetries of the Problem}

The soliton solution $\Phi= \la \Phi_0$ with $V^{\prime}(\la)=0$
 corresponds  to the  
projection operator $P_0$ onto the ground state of a one-dimensional 
harmonic oscillator
\beq
\label{proj}
\Phi_0 \sim P_0 \, = \, |0 \rangle \langle 0|
\end{equation}  
and is not the unique  solution to the equation $\Phi\s\Phi=\Phi$ (or,
in terms of the generic projector operator $P$, $P^2=P$).
Other solutions may be  obtained 
by choosing other projection operators \cite{stro} 
\beq
\Phi_n \sim P_n \, = \,  |n\rangle\langle n|, 
\end{equation}
or we can choose a superposition of solutions 
(a level $k$ solution in the terminology of \cite{stro}) 
\beq
\label{level}
\Phi_k \, = \, \la(\Phi_0 \, + \, \Phi_1 + \ldots \, + \, \Phi_{k-1}).
\end{equation}
Since the $\phi_m=|{m}\rangle\langle{m}|$ are a complete set
of projection operators, in the limit $k\rightarrow\infty$
 $\Phi_\infty=\la{\bf 1}$.
The corresponding soliton  takes the value $\Phi=\la$ everywhere, and
no spatial structure is present in $\mathbb{R}^2_\s$.

In the limit of infinite  non-commutativity the action (\ref{action})
 can be written 
in the operator form  
\beq
S \, = \, \int d^4x ~{\rm Tr}~\left[\ha \partial_a \hat{\Phi}\partial^a \hat{\Phi} - 
V(\hat{\Phi})\right],
\end{equation}
where $x^{a}$ are the commutative directions.    
 The action is manifestly invariant under a global 
$U(\infty)$ symmetry  
\beq
\hat{\Phi} \rightarrow U \hat{\Phi} U^\dagger.
\end{equation} 
In other words, if $\hat{\Phi}$
is a solution to the equations of motion -- in the limit of
large non-commutativity -- 
so is $U \hat{\Phi} U^{\dagger}$, where $U$ is any unitary
operator acting on ${\cal H}$.
A general Hermitian
operator $ \hat{\Phi}$ (corresponding to a real field $\Phi$)
may be obtained by acting on a diagonal operator (which 
corresponds to a radially symmetric field configuration)
by an element of the $U(\infty)$ symmetry group.
 Thus every solution to $V^\prime(\Phi)=0$
may be obtained from a radially symmetric solution
by means of $U(\infty)$ symmetry transformations and the 
solutions  consist of disjoint infinite dimensional
manifolds labeled by the set of eigenvalues of the corresponding operator.

The $U(\infty)$ symmetry is broken by any non-vanishing solution of
$\Phi(r)$ of Eq. (\ref{ext}). Consequently, every non-vanishing solution
implies a number of exact massless Goldstone modes corresponding
to
small displacements about $\Phi(r)$ on the manifold of solutions.
The generators of $U(\infty)$
are given by 
\beq 
R_{nm} \, = \, |n\rangle \langle m|\, + \, |m \rangle \langle n|
\end{equation}
 and
\beq
S_{nm} \, = \, i(|n\rangle \langle m|\, - \, |m \rangle \langle n|).
\end{equation} 
The Goldstone  modes are therefore given by the nonzero elements of
\beq
\delta \Phi \propto [R_{nm}, \, \Phi] \, , \quad  [S_{nm}, \, \Phi] \quad .
\end{equation}
Since in the   operator language the complex coordinates 
$z \, = \, (x^4 + i x^5)/ \sqrt{2}$ and 
$\bar{z} \, = \, (x^4 - i x^5)/\sqrt{2}$
correspond respectively to the annihilation operator
$a=z/\sqrt{\Theta}$ and destruction operator $\bar{a}=\bar{z}/\sqrt{\Theta}$,
and
\beq
[a,\bar{a}] \, = \, 1,
\end{equation}
the generator of the translations along the
coordinate $z$ ($\bar{z}$) is the operator $\bar{a}$ ($a$), so that
derivatives become
\beq
\partial_z = -\Theta^{-1/2}[\bar{a}, \cdot],~~~ 
\partial_{\bar{z}}=\Theta^{-1/2}[a,\cdot].
\end{equation}
Since 
\begin{eqnarray}
\left[R_{nm},|0\rangle \langle 0|\right]&=&\delta_{n0}\left[
|m\rangle \langle 0|-|0\rangle \langle m|\right]\nonumber\\
&-& \delta_{m0}\left[
|0\rangle \langle n|-|n\rangle \langle 0|\right],
\end{eqnarray}
and
\begin{eqnarray}
\left[S_{nm},|0\rangle \langle 0|\right]&=&i\delta_{n0}\left[
|m\rangle \langle 0|+|0\rangle \langle m|\right]\nonumber\\
&-& i\delta_{m0}\left[
|0\rangle \langle n|+|n\rangle \langle 0|\right],
\end{eqnarray}
the unbroken generators in the $\Phi_0$-background are given by $R_{00}$
and all the $R_{nm}$ and $S_{nm}$ with $m,n\geq 1$. 
Notice in particular that the  linear combination $\sum_{n\geq 0} 
(2n+1) R_{nn}$
reproduces the generator $(a \bar{a}+\bar{a}a)$ of rotations
in $\mathbb{R}_\s^2$. On the contrary, the generators $R_{0n}$ and $S_{n0}$
with $n\geq 1$ are broken in the $\Phi_0$ background and give rise
to an infinite tower of Goldstone modes $\left\{\varphi_{0n},\varphi_{n0}
\right\}$.
The soliton solution $\Phi_0$ therefore breaks the global
symmetry $U(\infty)$ down to 
$U(\infty-1)\otimes U(1)$.

All these considerations refer to the case of infinite non-commutativity.
At finite $\Theta$, the
kinetic term in action (\ref{action}) explicitly breaks the
$U(\infty)$ symmetry
down to $ISO(2)$, the Euclidean  group in the two extra
dimensions.
Finite $\Theta$ effects will therefore 
lift the $\Theta\rightarrow \infty$ manifold of solutions to
a discrete set of solutions, still guaranteeing  the existence of
at least one classically stable soliton, which is identifiable
with the  the Gaussian 
$\la\Phi_0(r^2)$ \cite{stro}. Finite $\Theta$ effects are also expected
to  provide  masses to the (unwanted) 
Goldstone modes $\left\{\varphi_{0n},\varphi_{n0}
\right\}$. Alternatively, one can promote the $U(\infty)$ symmetry to a gauge
symmetry with the components of the gauge fields  
transverse to the soliton behaving as  
adjoint scalar fields on the soliton \cite{harvey}. As we shall see,  
this  gauge symmetry removes 
from the spectrum the Goldstone bosons  of the soliton 
through the Higgs mechanism, leaving
behind a massless gauge field in the 4D action.

Our strategy is therefore the following. We will work in the
limit 1 TeV $\ll 1/\sqrt{\Theta}\ll M_f$, where $M_f$ is the fundamental
scale of the 6D theory. This range of $\Theta$ allows the presence of 
the stable soliton $\Phi_0$ and at the same time  
to treat perturbatively 
   the kinetic terms
along the non-commutative directions  in the action of  
matter fields and gauge fields coupled to
the soliton. 

\section{Fluctuations around the soliton}
In the rescaled coordinates the action of the real
scalar field $\Phi$ is given by
\beq
S[\Phi] \, = \, \int d^6x \, \Theta \left[ \ha \de_a \Phi  \de^a \Phi 
\, + \, \Theta^{-1} \ha \de_i \Phi \star 
\de^i  \Phi \, - \, V(\Phi)
\right] \quad .
\end{equation}
Let $\la$ be a real root of $V^\prime(\la) = 0$. 
In the limit of large $\Theta$ a non trivial static
solution is $\Phi = \la \Phi_0\sim |0\rangle\langle 0|$,
 with $\Phi_0 \star \Phi_0 = \Phi_0$. 

To find the 4D action, we  apply some of the powerful simplifications
following from non-commutative geometry.
Using the operator correspondence, a complete set of functions
in $\mathbb{R}_\s^2$ is given by 
\beq
\Phi_{mn}(x^i)\sim |m\rangle\langle n|,
\end{equation}
so the general fluctuation around the soliton $\Phi_0$ is
\beq
\label{fl}
\Phi = \la  \Phi_0(x^i) \, + \, \varphi \, , \qquad \varphi \, = 
\, \sum_{m,n=0}  \varphi_{mn}(x^a) \Phi_{m n}(x^i)
\quad .
\end{equation}
Consider $V(\Phi)$ of the form $V(\Phi) = c_2 \Phi^2 - c_3  \Phi^3 +  
c_4  \Phi^4$, where all $c_i$'s are positive and we do not allow
a linear term (otherwise $\Phi=\la\Phi_0$ cannot be a solution
of Eq.(\ref{ext})).
The extremum
condition reads $2 c_2 \la - 3 c_3 \phi_*^2 + 4 c_4 \phi_*^3 = 0$.
Suppose the $c_i$'s are such that $V^\prime(\la)=0$, $V^{\prime\prime}
(\la)>0$ for some non-vanishing $\la$. Performing the traces,
one gets
\bea
&&
{\rm Tr} \left[( \la  \Phi_0 \, + \, \varphi)^2 \right] \,  = \, \phi_*^2
 \, + \, 2 \la 
\varphi_{00} \, + \,   \sum_{m,n}  \varphi_{mn} \, \varphi_{nm} \quad , \nb \\
&&{\rm Tr} \left[( \la  \Phi_0 \, + \, \varphi)^3 \right] \,  = \, 
3 \la \sum_m  \varphi_{m0} \, \varphi_{0m} \, + \, 3 \phi_*^2
 \varphi_{00} \, + \,\phi_*^3 \, + \, {\cal O}(\varphi^3) \; ,
\\   
&& {\rm Tr} \left[( \la  \Phi_0 \, + \, \varphi)^4 \right] \,  = \,
4 \phi_*^2 \sum_m  \varphi_{m0} 
\, \varphi_{0m} \, + \, 2 \phi_*^2 \varphi_{00} \, + \, 4 
\phi_*^3 \varphi_{00} \, + \, \phi_*^4 \, + \, {\cal O}(\varphi^3)  . \nb
\ena
Making use of the operator correspondence (\ref{cor}) and the fact 
that $\Phi_0\sim |0\rangle\langle 0|$ is orthogonal
to $\Phi_{mn}$ for $m,n>0$, 
 one gets  the following 4D effective action
\bea
S[\la \Phi_0 \,+ \, \varphi] \,&& = \pi\,\Theta\, \int d^4x  \left
 \{\ha \de_a \varphi_{00}
\, \de^a \varphi_{00} \, - \, \ha V^{\prime \prime}(\la) \varphi_{00}^2 \, + \, V(\la) \right.  
\nb \\
&&\ + \sum_{m>0, n>0}
\left( \ha \de_a \varphi_{mn} \, \de^a \varphi_{nm} \, - \, c_2 \varphi_{mn} \, \varphi_{nm} \right)
 \nb \\
&&\left. + \, \sum_{m>0} \left[\de_a \varphi_{m0} \de^a \varphi_{0m} 
\, - \, \left(2 c_2 \, - \,  3 
\la c_3 \, + \, 4 c_4 \phi_*^2 \right) \, 
 \varphi_{m0} \, \varphi_{0m} \right] \right \}  .
\ena
Taking into account that $\la$ is an extremum, we are left with 
\bea
\label{mass}
S[\la \Phi_0 \,+ \, \varphi] \,&& =  \pi\,\Theta\,
 \int d^4x \left [\ha \de_a \varphi_{00}
\, \de^a \varphi_{00} \, - \, \ha V^{\prime \prime}(\la) \varphi_{00}^2 \, + \, V(\la) \right.  
 \\
&&\left. + \sum_{m>0, n>0}
\left( \ha \de_a \varphi_{mn} \, \de^a \varphi_{nm} \, - \, c_2 \varphi_{mn} \, \varphi_{nm} \right)
+ \, \sum_{m>0} \de_a \varphi_{m0} \de^a \varphi_{0m}  \right]  \;   . \nb
\ena
The four-dimensional action is therefore given by the heavy
mode $\varphi_{00}$ with mass squared $V^{\prime\prime}
(\la)$ (of the order of $M_f^2$), by the tower of heavy states 
$\varphi_{nm}$ with mass $2c_2$ (again of the order of $M_f^2$), and by the
announced  tower of Goldstone modes $\left\{\varphi_{0n},\varphi_{n0}
\right\}$ (notice that $\varphi_{0n}=\varphi^*_{n0}$ because of the
reality of the field $\Phi$). We will see in the following how these
unwanted Goldstone modes disappear from the 4D effective action
once the $U(\infty)$ symmetry is gauged or -- alternatively --
when finite $\Theta$-effects are taken into account\footnote{The analysis can 
be easily repeated with a more general solution of Eq.
(\ref{ext}). In general, the soliton solution will contain operators of the
form $|m \rangle \langle n|$, $m\neq n$, corresponding to non-spherically
symmetric soliton fluctuations in the four-dimensional action.}.

\section{Localization of scalar fields}
Let us now consider a real scalar field $\chi$ coupled to the 
non-commutative soliton $\Phi_0$ in the 6D theory. 
We will  show that -- according to the 
type of interaction we allow for  the field  $\chi$ 
with the soliton -- different
four-dimensional spectra may arise. In particular, we may obtain 
a four-dimensional
action containing: either  {\it i)} a zero mode together with a tower of 
{\it degenerate} heavy states, or {\it ii)} a zero mode with a tower of
Kaluza-Klein heavy states, or {\it iii)} only a zero mode!
Let us see how this comes about.

\subsection{The four-dimensional action with a zero mode and  a tower of 
degenerate heavy states}
Let us consider the following action for the scalar field $\chi$
\bea
&&S =  \int d^6x  \left[ \ha   f(\Phi) \star \de_\mu \chi  \star \de^\mu \chi \, - \,
 \ha  m^2(\Phi) \star \chi\star g(\phi_\ast-\Phi)
 \star \chi \right]  \label{nce1}\\
&&=  \int d^4x d^2z \left \{ \ha f(\Phi)  \star \left[ \de_a \chi \star
\de^a \chi \, - \, \partial_z \chi \star 
\partial_{\bar{z}} \chi \, - \, \partial_{\bar{z}} \chi \star
\partial_z \chi \right] \right. \nb \\ 
&& \left. - \, \ha  \, m^2(\Phi) 
\star \chi \star g(\phi_\ast-\Phi) \star \chi \right\} . \nb 
\ena 
We stress that in (\ref{nce1}) -- due to the lack of 
commutativity -- the derivatives 
$\partial_z \chi \star \partial_{\bar{z}} \chi$
and $\partial_{\bar{z}} \chi \star \partial_z \chi$ do not sum up.
As for the fluctuations around the soliton, we can  expand the field
$\chi$ as
\beq
\label{expansion}
\chi(x^\mu)= \sum_{m,n\geq 0}  \chi_{m n}(x^a) \Phi_{mn}(x^i) 
\quad .
\end{equation}
The 4D effective action is obtained integrating out 
the   non-commutative coordinates and setting
$\Phi = \la \Phi_0$. In the limit of infinite non-commutative parameter, 
the  kinetic term from the extra dimensions 
is negligible.
Since  $\Phi_0$ is a projector
operator, when $\Phi = \phi_\ast \Phi_0$, we may write the crucial relation
\beq
g(\phi_\ast -\Phi) \, = \, g(\phi_\ast) \, (1-\Phi_0).
\end{equation}
This will have the effect of projecting out the zero mode
$\chi_{00}$ from the mass term. Indeed -- integrating 
over the two extra-dimensions -- the 4D action reads
\bea
&&S \, = \,  \pi \, \Theta \,\int d^4x  \left \{\ha f(\la) \de_a \chi_{00}  
\, \de^a \chi_{00} \,+\, \sum_{n \geq 1}  \left[
\ha f(\la) \, \de_a \chi_{0n}  \, \de_a \chi_{n0} \right. \right. \nb \\
&&\qquad \left.  \left. - \, \ha m^2(\la)g(\phi_\ast) \, \chi_{n0}
 \chi_{0n} \right]
\right \} \quad .
\ena
The fluctuations
$\chi_{m n}(x^a)$ with $m,m\geq 1$ have been projected out
since they are orthogonal to the soliton $\Phi_0\sim |0\rangle\langle 0|$.
This property makes us the favour
of rendering  the $U(\infty-1)$ part of the 
unbroken symmetry  irrelevant as far as the four-dimensional
physics is concerned.
Furthermore, the modes $ \chi_{n0}$ with $n\geq 1$ show up with 
 the same 
terms in the limit of infinite non-commutativity. This is due to the 
residual $U(\infty-1)\otimes U(1)$ symmetry left unbroken by the
soliton (it is easy to see that the generators $R_{00}$ and $R_{mn}$, $S_{mn}$
with $m,n\geq 1$ leave the term  $m^2(\Phi)
\star \chi \star g(\phi_\ast - \Phi)\star\chi$ invariant
upon integration over the two non-commutative directions).
The 4D spectrum contains a real zero mode $\chi_{00}$ plus a tower
of (arbitrarily) massive complex states $\chi_{n0}$ ($n\geq 1$) 
which are degenerate 
in the limit of infinite 
non-commutative parameter.  
It is also remarkable that the localization of the massless
mode $\chi_{00}(x^a)$ is due entirely to its kinetic term. 
This originates from the property  (\ref{crucial}): any function
$f(\Phi)$ in front of the kinetic term is proportional to
the Gaussian soliton $\phi_0$ which acts as a warp-factor.
Furthermore,  the 4D action inherits informations about the non-commutative
soliton only in  the point $\Phi=\la$. No detailed knowledge of the
function $f(\Phi)$ is needed and the 4D action for the
massless mode could be rewritten as 
\beq
S \, = \, \int d^4x \, d^2z \,\delta^{(2)}(z-z_*)\, 
f(\Phi)\, \frac{1}{2} \, \partial_a\chi_{00} \, \partial^a\chi_{00},
\end{equation}
where $z_*$ is the point in the $\mathbb{R}_\s^2$ space where the soliton
$\Phi_0$ takes the value $\phi_0=\la$.
This is quite surprising if we think that 
the soliton spreads in the $\mathbb{R}^2_\s$ over 
distances $\sim\sqrt{\Theta}\gg M_f^{-1}$. 
We believe this is another manifestation of the UV/IR
connection pointed out first in \cite{con}.

\subsection{The four-dimensional action with a zero mode and  a tower of 
Kaluza-Klein states}
Let us now take  the following simple 
quadratic action for the field $\chi$
\bea
&&S \, = \, \int d^4x \, d^2z \left( \ha   f(\Phi) \star 
\de_\mu \chi  \star \de^\mu \chi \, \right)  \label{nce}\\
&&= \, \int d^4x \, d^2z \left[ \ha f(\Phi)  \star \left( \de_a \chi \star
\de^a \chi \, - \, \partial_z \chi \star \partial_{\bar{z}} \chi \, - 
\, \partial_{\bar{z}} \chi \star
\partial_z \chi \right) \right] . \nb 
\ena 
Performing again  the expansion (\ref{expansion}) and neglecting the
finite $\Theta$-effects, the 
4D action reads
\beq
S \, = \pi \, \Theta \,\int d^4x  \left(\ha f(\la) \de_a \chi_{00}  
\, \de^a \chi_{00} \, + \, \ha \sum_{n \geq 1}  f(\la) \, \de_a \chi_{0n}  \,
 \de_a \chi_{n0} 
\right) \quad .
\end{equation}
The  finite $\Theta$ 
correction comes from the kinetic ${\cal L}_{K}$ term containing the derivatives 
along the non-commutative coordinates. The latter 
can be  written as (in the rescaled coordinates)
\bea
&&{\cal L}_{K} \, = \, - \frac{f(\la)} {2 \Theta}\int d^2w \, \ha \, \Phi_0
 \star \left(\partial_w \chi \star \partial_{\bar{w}} \chi \, + \, \partial_{\bar{w}} \chi  \star 
\partial_w  \chi\right)\,  =  \\ 
&&= \, -\frac{\pi}{\Theta} f(\la)
\, \, \ha \sum_{m,n,r,s \geq 0} \chi_{mn} \, \chi_{rs} \, 
{\rm Tr}~ \left \{ \Phi_0 \, 
\left[a, \, \Phi_{mn}  \right] \, \left[\Phi_{rs}, \, \bar{a} \right] \, + \, \Phi_0 \, 
\left[\Phi_{rs}, \, \bar{a} \right] \, \left[a, \, \Phi_{mn}  \right]   \right \}. \nb 
\ena
Using the relations 
\bea
&&\left[a, \, \Phi_{mn} \right] \, = \, \sqrt{m} \, \Phi_{m-1 \, n} \, - \, \sqrt{n+1} \, \Phi_{m \, n+1},
\nb \\
&&\left[\Phi_{rs}, \,  \bar{a} \right] \, = \, \sqrt{s} \, 
\Phi_{r \, s-1} \, - \, \sqrt{r+1} \, 
\Phi_{r+1 \, s} \quad , 
\ena
we find the kinetic term ${\cal L}_{K}$ 
\begin{eqnarray}
\label{nn}
{\cal L}_{K}  &=&   
- \frac{\pi f(\la)}{\Theta} \left[\left(\chi_{00}- \chi_{11}\right)^2+
4\chi_{10}\chi_{01}-\sqrt{2}\chi_{01}\chi_{21}-\sqrt{2}\chi_{10}\chi_{12}+
\chi_{12}\chi_{21}
\right. \\
&+&\left.\sum_{n\geq 2}(2n+1)\chi_{n0}\chi_{0n}
-\sum_{n\geq 3}\sqrt{n}\chi_{0,n-1}\chi_{n1}
-\sum_{n\geq 3}\sqrt{n}\chi_{1n}\chi_{n-1,0}+
\sum_{n\geq 3}\chi_{1,n}\chi_{n1}\right]. \nb 
\end{eqnarray}
It is clear from (\ref{nn}) that the 
degrees of freedom associated with $\chi_{1 n}$ with $n\geq 1$ do not  
propagate, indeed the corresponding equations of notions lead to the following algebraic constraints
\beq
\chi_{1n} \, = \,  \sqrt{n} \,  \chi_{0 n-1} \, ,  \; \; n \geq 1 \quad .
\end{equation}
The effective action for the propagating degrees of freedom takes the form
\bea
&&S \, =  \pi  \, \Theta \, \int d^4x  \left \{ \ha f(\la) \de_a \chi_{00}  
\, \de^a \chi_{00} \, + \, \sum_{n \geq 1} \left[ \ha f(\la) \de_a \chi_{0n}  \, \de^a \chi_{n0} \nb 
\right. \right. \nb \\ 
&& \left. \left. \qquad  - \, 
\ha      f(\la)(n+1) \Theta^{-1}  \chi_{0n}  \chi_{n0} \right] \right \}.
\label{eff2} 
\ena
The kinetic term ${\cal L}_{K}$ provides all the modes 
$\chi_{0n}(x^a)$ with $n\geq 1$ with an extra mass squared  
$\sim (n+1)f(\la)/\Theta$.
Therefore, 
the four-dimensional spectrum contains\footnote{A similar computation shows
that the same result applies to the massless modes $\varphi_{0n}(x^a)$
of the fluctuations of the soliton in the case in which the
kinetic term is multiplied by a function $f(\Phi)$ (and the $U(\infty)$ 
symmetry is not gauged). This is equivalent to say that
the soliton $\Phi_0$ is stable against small fluctuations around it.
Of course, the modes $\varphi_{01}$ and
its complex conjugate $\varphi_{10}$ remain massless since they
represent the translational zero modes 
arising from the spontaneous  breaking of
translation invariance
along the directions $x^4$ and $x^5$.}
 a massless real 
state $\chi_{00}(x^a)$ and
an infinite tower of massive  complex  modes $\chi_{0n}(x^a)$ 
with mass squared and spacing proportional   to $1/\Theta\gg {\rm TeV}^2$! 
The four-dimensional action
therefore looks like a 5D Kaluza-Klein theory compactified on a circle
of radius $\sqrt{\Theta}$. This  phenomenon  arises from the
fact that 
the degrees of freedom propagating in the 
two extra dimensions, due to the commutation relations, 
actually  correspond to a point-particle in one dimension. 
This is a remarkable result. Not only the presence of 
the soliton $\Phi_0$ projects out
all the modes $\chi_{mn}$ with $m,n\geq 1$ from the 4D effective theory,
but also leaves behind a single massless mode plus
a tower of very heavy KK-states. This happens in spite
of the fact that the two extra dimensions we started from are infinite.
Again,   the localization of the massless
mode $\chi_{00}(x^a)$ is due entirely to its kinetic term.

\subsection{The four-dimensional action with the zero mode only}
Consider  now the following action for the scalar field $\chi$
\beq
S \, = \, \int d^4x \, d^2z \, \ha   f(\Phi) \star \de_\mu \chi  \star
g(\Phi)\star \de^\mu \chi  \, .
\end{equation}
In the infinite limit of the non-commutative parameter, the four-dimensional
action reads
\beq
S \, = \pi \, \Theta \,\int d^4x  \,\ha f(\la) g(\la)\de_a \chi_{00}  
\, \de^a \chi_{00} \, .
\end{equation}
It is easy to
see that the kinetic term ${\cal L}_{K}$  
depending on the non-commutative coordinates gives rise to a 4D action
of the type $\int d^4x \left|\chi_{01}\right|^2$ and -- since the fields
$\chi_{01}$ and $\chi_{10}$ do not propagate -- it vanishes identically.
Therefore, all modes but the zero mode $\chi_{00}$ have been projected out! 
The four-dimensional action has lost completely the notion of being
immersed in the extra-dimensional world.

\section{Localization of chiral fermions}
In this section we sketch how to 
extend  the previous analysis to fermions. 
The Dirac matrices which generate the Clifford algebra are taken to be
\beq
\Gamma^a \, = \, \begin{pmatrix} \gamma^a & 0   \\   0  &  - \gamma^a \end{pmatrix}, 
\qquad  \Gamma^4 \, = \, i \, \begin{pmatrix}   0 &   \mathbf{1} \\    \mathbf{1} &  0  
\end{pmatrix}, \qquad  \Gamma^5 \, = \,  \begin{pmatrix} 0 &  \mathbf{1} \\ - \mathbf{1} 
& 0 \end{pmatrix}  \quad .
\label{gamma}
\end{equation} 
We define the chirality operator
\beq
\overline{\Gamma} \, = \, \Gamma^0 \, \Gamma^1 \, \cdots \, \Gamma^5 \, = \, \begin{pmatrix} - \gamma^5 &
0 \\ 0 & \gamma^5 \end{pmatrix} \quad ,
\end{equation}
which has the properties
\beq
\left(\overline{\Gamma}\right)^2=1,~~~~ \left\{\overline{\Gamma},
\Gamma^\mu \right\}=0\, .
\end{equation} 
A Dirac spinor in six dimensions has 8 complex components and  
can decomposed as the direct sum of a pair 
of 6D Weyl spinors
\beq
\Psi \, = \, \begin{pmatrix} \zeta \\ \psi \end{pmatrix} \quad ,
\label{spin}
\end{equation}
where each Weyl spinor  corresponds to a Dirac spinor in 4D. One can also
define two 6D Weyl spinors
\beq
\Psi_-=\frac{1-\overline{\Gamma}}{2}\Psi=
\begin{pmatrix} \zeta_R \\ \psi_L \end{pmatrix}\, , ~~~~~ 
\Psi_+=\frac{1+\overline{\Gamma}}{2}\Psi=
\begin{pmatrix} \zeta_L \\ \psi_R \end{pmatrix}\,  ,
\end{equation}
each of them representing a Dirac spinor in 4D. 
Each spinor can be decomposed as usual as
\beq
\label{dec}
\zeta(x^\mu) \, = \, \sum_{m, n  \geq 0} \zeta(x^a) \, \Phi_{mn}(x^i) \, , 
\qquad 
\psi(x^\mu) \, = \, \sum_{m, n  \geq 0} \psi(x^a) \, \Phi_{mn}(x^i)  \quad . 
\end{equation}
Since the generator of the isometry group $U(1)$ which is
left unbroken by the
soliton $\Phi_0$ is given in polar coordinates $(r,\theta)$ by
$\left(\partial_\theta+\ha i\tau_3\right)$ \footnote{Notice that the
commutation relations between $z$ and $\overline{z}$ are invariant
under translations and rotations and spinors have definite
properties under these transformations.}
and  
$i\partial_\theta$ in the operator language is proportional to 
$(\overline{a}a+a\overline{a})/2$, 
it easy to see that the fermions $\zeta_{n0}$ and $\zeta_{0n}$ have charge
$n+\ha$ and $-n+\ha$ respectively 
 under the isometry group $U(1)$, while   the fermions 
$\psi_{n0}$ and $\psi_{0n}$ have  charge
$n-\ha$ and $-n-\ha$, respectively. 

For the sake of simplicity, we consider the following action in the
limit of infinite non-commutativity
\bea
&&S \, = \, \int d^6x \left[ \frac{i}{2} \, f(\Phi) \star \left(
 \overline{\Psi} \star g(\Phi)\star \Gamma^\mu \de_\mu \Psi \, - \, 
 \de_\mu \,  \overline{\Psi} \star g(\Phi)\star \Gamma^\mu   \, \Psi \right) \,
\right] \nb \\
&&\qquad \qquad = \, \int d^6x \left( \frac{i}{2} \, f(\Phi) \star  
\overline{\Psi} \star g(\Phi)\star  
\stackrel{\leftrightarrow}\de_\mu \Gamma^\mu  \Psi  \right) \quad .
\label{fa1}
\ena
The 4D action reads
\beq
S \, = \, \Theta \pi \int d^4x \,       f(\phi_*)\,g(\phi_*)  
\left( \overline{\psi}_{00} \gamma^a 
\stackrel{\leftrightarrow}{\de_a} \psi_{00} \, + \, \overline{\zeta}_{00}
 \gamma^a 
\stackrel{\leftrightarrow}{\de_a} \zeta_{00} \right).
\end{equation}
We are left with a  four-dimensional action containing two 4D Dirac 
massless spinors whose  localization is due once more
to its kinetic term and to the role of warp factor played by the soliton.
The relevant point though is that -- had we started
from a Weyl spinor in 6D  -- we would have
ended up with a chiral theory! Indeed, suppose we start with only the
Weyl spinor $\Psi_-$ ($\Psi_+$). The 4D theory then contains the
two Weyl 4D spinors $\psi^{00}_L$ and  $\zeta^{00}_R$ 
($\psi^{00}_R$ and  $\zeta^{00}_L$). Since 
$\psi^{00}_L$ ($\psi^{00}_R$) is a  
 spinor with charge $-\ha$ and  $\zeta^{00}_R$ ($\zeta^{00}_L$) its
with charge $\ha$, the theory
admits chiral fermions with respect to the isometry group $U(1)$ which is left
unbroken by the presence of the noncommutative soliton. 

Of course and
in complete similarity with the scalar case analyzed in the previous section,
different choices of  the starting  action other than (\ref{fa1}) lead
to different fermionic spectra.

\section{Localization of gauge fields}

As we have seen in Section 3, the soft fluctuations around the soliton $\Phi=
\phi_*\Phi_0$
are made of a tower of massless states $\varphi_{0n}$ with $n\geq 1$ in 
the limit of infinite non-commutative parameter. They reflect the
symmetry breakdown of $U(\infty)$ symmetry down to $U(\infty-1)\otimes U(1)$
caused by the presence of the soliton $\Phi_0$ in $\mathbb{R}_\s^2$. 
These 
Goldstone are not necessarily problematic from the
four-dimensional point of view as 
they may acquire a mass through the  finite $\Theta$-effects induced
by the kinetic term along the two non-commutative directions. 

There is an alternative route one may take, though.
If the global  $U(\infty)$ symmetry is promoted 
to a gauge symmetry, the latter
becomes a $U(1)$ gauge symmetry of the non-commutative theory and 
removes 
from the spectrum the unwanted Goldstone bosons; they are simply  
eaten by the Higgs mechanism. 

This procedure has another advantage. It
leaves the four-dimensional effective action only with a   
massless $U(1)$ gauge field\footnote{For alternative higher-dimensional
mechanisms to localize gauge fields, see \cite{sh1,sh2}.}
(plus  the components of the gauge fields  
transverse to the soliton which behave as  
adjoint Higgs fields -- the graviphotons --  
on the soliton and play  the role of the Goldstone
modes  from the breaking of the translation invariance along the 
$(x^4,x^5)$-directions). 

Therefore, the presence of the radially symmetric
soliton $\Phi_0$ in  $\mathbb{R}_\s^2$ and the corresponding breaking of the 
 gauge symmetry  $U(\infty)$
provides a  quantum field-theoretic mechanism to localize a  $U(1)$ 
gauge field onto a four-dimensional brane starting from a six-dimensional
world with infinite extra-dimensions.
This does not come as a surprise as 
our procedure is the quantum field-theoretic analog
of the  one adopted
by Harvey et {\it al.} in Ref. \cite{harvey} where -- in a string-theoretic
context -- it was shown that the non-commutative geometry induced by a large
auxiliary magnetic field allows the identification of D-branes in open
string theory with the non-commutative tachyonic solitons.

\subsection{Promoting the global $U(\infty)$ symmetry to a gauge symmetry}
The gauge transformation law of a   
non-commutative  gauge field $A_\mu$ is given by 
\begin{equation}
\delta_\epsilon A_{\mu} = \partial_\mu \epsilon - i[A_\mu,\epsilon],  
\end{equation}
where 
\begin{equation}
[A_\mu,\epsilon] = A_\mu \star \epsilon - \epsilon \star A_\mu.
\end{equation}
The corresponding field strength is 
\begin{equation}
F_{\mu\nu} = \partial_\mu A_\nu - \partial_\nu A_\mu -  
i[A_\mu,A_\nu].
\end{equation}
When non-commutativity is present, even the transformation law for
an abelian $U(1)$ gauge field is nontrivial. This happens
because the commutator of two infinitesimal gauge transformations
with generators $\epsilon_1$ and $\epsilon_2$ a gauge transformation
generated by $i(\epsilon\star\epsilon_2-\epsilon_2\star\epsilon)$.
Such commutators are nontrivial in the case of a gauge group $U(N)$ with rank
1, {\it i.e.} $N=1$, even though in the limit $\Theta=0$ the rank 1 case is the
abelian $U(1)$ gauge group.

We now assume that the scalar field $\Phi$ transforms in the adjoint
of a non-commutative  $U(1)$: 
\beq
\delta_\epsilon \Phi = -i[\Phi,\epsilon],
\end{equation} 
so that the 
 covariant derivative reads
\beq
D_\mu \Phi = \partial_\mu \Phi -i[A_\mu,\Phi].
\end{equation} 
 
The 6D action is then 
\beq
S = \int \! d^{6}x \, \left[
\ha f(\Phi)D^\mu \Phi D_\mu \Phi - V(\Phi) - \frac{1}{4} h(\Phi)
F^{\mu\nu}F_{\mu\nu} 
\right]. 
\end{equation} 

Again, we take first the limit of infinite non-commutative parameter.
The derivatives along the non-commutative directions $x^i$ are
all suppressed, yielding  

\begin{eqnarray}
\label{gauge}
S &=& \int \! d^{6}x \,   
\left[ \ha f(\Phi)D^a \Phi D_a \Phi - \ha h(\Phi)D^a A_i D_a A^i 
- V(\Phi)\right.\nonumber \\
& -&\left. \ha f(\Phi)[A_i,\Phi][A^i,\Phi]   
 +\frac{1}{4} h(\Phi)[A_i,A_j][A^i,A^j]  
-\frac{1}{4} h(\Phi)F^{ab}F_{ab}  \right].  
\end{eqnarray}  
In the limit of infinite non-commutativity,  the 
$A_i$ $(i=4,5$) become adjoint scalar fields and the   action becomes 
invariant 
under the gauge transformations 
\begin{eqnarray}
\label{tr}
\delta_\lambda \Phi &=& -i[\Phi,\lambda], \nonumber\\ 
\delta_\lambda A_i &=& -i[A_i,\lambda],\nonumber\\
\delta_\lambda A_a &=&\partial_a \lambda -i[A_a,\lambda].  
\end{eqnarray}
Using now the correspondence (\ref{cor}), the action (\ref{gauge}) appears
from the four-dimensional point of view as the action of a $U(\infty)$
gauge theory coupled to the adjoint scalars $\Phi$ and $A_i$ (for instance,
the infinitesimal transformation (\ref{tr}) of the field $\Phi$ 
is the infinitesimal form of the transformation
$\hat{\Phi} \rightarrow U \hat{\Phi} U^\dagger$). The gauge symmetry remains 
exact even for finite non-commutativity \cite{harvey} provided that 
the derivatives $\partial_i\epsilon$ reappear in $\delta_\epsilon A_i$.

\subsection{The fluctuations of the gauge fields around the soliton}
Since the action (\ref{gauge}) is invariant under a gauge $U(\infty)$ symmetry,
we make use of the gauge freedom to work in the unitary gauge along which
the fluctuations around the soliton (\ref{fl}) are such that  
\beq
\varphi_{0n}=\varphi_{n0}\, =\, 0 ~~~{\rm for}~~~ n\geq 1.
\end{equation}
The  fluctuations of the gauge fields can be decomposed as usual as  

\begin{eqnarray}
A_a(x^\mu) &=&  
\sum_{m,n\geq 0}A_a^{mn}(x^a) \Phi_{mn}(x^i),\nonumber\\ 
A_i(x^\mu) &=& \sum_{m,n\geq 0}A_i^{mn}(x^a) \Phi_{mn}(x^i),  
\end{eqnarray}
where $A_a^{mn}$ and $A_i^{mn}$ represent  Hermitian  matrices. 

Plugging these
fluctuations into the action (\ref{gauge}) and computing the
trace,  we find that all modes 
with $m,n\geq 1 $ are again projected out by the soliton background
at the quadratic order. 
What is left in the spectrum are the fields  
$\varphi_{00}$, $A_a^{00}$, $A_i^{00}$, together with  the $0m$- and 
$m0$-components of fields $A_i$ and  $A_a$. 

The action for the $00$-components of the fields is given by 
\beq
\label{4d}
S =\pi\,\Theta \int \!  d^{4}x  \left[ 
\ha f(\phi_*) \partial^a \varphi_{00} \partial_a \varphi_{00} - 
\ha V^{\prime\prime}(\phi_*)\, \varphi^2_{00}
+ \ha h(\phi_*) \partial^a  A_i^{00} \partial_a  A_i^{00} 
- \frac{1}{4} h(\phi_*) F^{ab}F_{ab}\right], 
\end{equation}
where $F_{ab} = \partial_a A^{00}_b - \partial_b A^{00}_a$ is the standard  
field strength.  This is the  4D action describing a massless $U(1)$
gauge field plus the two transverse  
scalars $A_i^{00}$ which play the role of the  Goldstone modes
arising from the spontaneous breakdown of the translation invariance along the
$(x^4,x^5)$-directions.
 It is easy to understand why the
$U(1)$ gauge field $A_a^{00}$ remains massless; 
there is simply no Goldstone mode
$\varphi_{00}$ to be eaten. The  latter 
is indeed massive, see Eq. (\ref{mass}).
Another way of thinking of it is that  $A_a^{00}$ is the massless 
gauge field
corresponding  to the residual  $U(1)$ symmetry (or rotational
isometry) which is left unbroken in the spontaneous breakdown process
$U(\infty)\rightarrow U(\infty-1)\otimes U(1)$. Furthermore, the fact that
the symmetry is a gauge symmetry guarantees that the field $A_a^{00}$
remains massless even when finite $\Theta$-effects are taken into account
\footnote{Similarly, 
had we started from a level $k$ soliton solution (\ref{level}),
the resulting unbroken gauge symmetry would have been the non-Abelian 
$U(k)$ symmetry.}. 

To see explicitly how the action (\ref{4d}) comes about, take for example the 
term 
\beq
\int \!  d^{6}x  f(\Phi)[A_\mu,\Phi][A^\mu,\Phi]
\end{equation}
 present in the  action  (\ref{gauge}). 
Using the relation $ f(\phi_*\Phi_0)=f(\phi_*)\Phi_0$
and carrying out  the trace, it gives rise to
\beq
\int \!  d^{4}x  f(\phi_*)\phi_*^2
\left(A_\mu^{00}A^\mu_{00}-\sum_{n\geq 0}
A_\mu^{0n}A^\mu_{n0}
\right)=-\sum_{n\geq 1}\int \!  d^{4}x  f(\phi_*)\phi_*^2~A_\mu^{0n}A^\mu_{n0}.
\end{equation}
This shows explicitly how the modes $A_a^{00}$, $A_i^{00}$ remain massless.
On the other hand,  the $0m$ and $m0$ 
components of fields $A_\mu$ acquire a mass  
\beq
m_A^2 = 
\frac{f(\phi_*)}{h(\phi_*)}\, \phi_*^2
\end{equation}
being the 
corresponding quadratic 
action 
\beq 
S =  \pi\, \Theta\,\sum_{n\geq 1}  
\int \! d^{4}x \, \left[ 
\ha h(\phi_*)\partial^a A_i^{0m} \partial_a A^i_{m0} 
- \ha f(\phi_*)\phi_*^2 A_i^{0m} A^i_{m0} \right]. 
\end{equation}
This is the announced Higgs mechanism. Thus, we find the remarkable
result that -- starting from a six dimensional theory with infinite
extra dimensions -- the non-commutative soliton projects out all the
undesired gauge field states leaving behind only one massless
photon (plus the graviphotons) whose localization on the soliton
is only due to its kinetic term.

\section{Conclusions and perspectives}
In this paper we have examined the construction of what  we dubbed the
non-commutative world brane. The underlying
theory is 
a six-dimensional non-commutative field theory where  non-commutativity
shows up            along  the two extra infinite spatial dimensions. The  
four-dimensional world-volume  is  
a scalar  soliton living in the non-commutative space. It appears that
such a construction provides a mechanism to localize scalar, fermionic and
gauge fields. Let us now consider the advantages and the possible
obstacles to such a mechanism. 

First, the advantages: the 
study of localization of fields inside the non-commutative scalar soliton
is made easy by the powerful correspondence between non-commutative 
coordinates and operators on a single particle Hilbert space. 
Equations of motion in field theory translate into 
algebraic operator equations and integrals over the two extra dimensions
reduce to simple traces over operators expressed in terms of the
eigenstates of a simple one-dimensional oscillator system. Under this
correspondence, the non-commutative soliton behaves like a projector
operator and deriving the spectrum of the low-energy four-dimensional
action is rather simplified. Field localization takes place
through the coupling to  the soliton in the kinetic term and -- thanks to
the crucial property (\ref{crucial}) -- the soliton acts like a warp
factor which is quite insensitive to  the form of the coupling itself.
Yet changing the latter gives rise to  
diverse  low-energy spectra. On a more phenomenological side, our
construction  admits chiral fermionic zero modes
and it ensures the localization of  massless gauge fields 
 very much the same as it happens in string theory 
where 
gauge fields (interpreted as the end-points of open strings) get
localized on D-branes identified with non-commutative tachyonic solitons
when a large
auxiliary magnetic field is turned on.

Now the possible obstacles. At present, we do not know how gravity
behaves in the presence of the non-commutative soliton and if 
four-dimensional observers may still empirically
deduce a four-dimensional Newton's law. In fact, the study of
gravity in non-commutative spaces is just as its infancy. In
 non-commutative spaces the metric becomes complex and one 
can obtain complexified gravity in $D$ dimensions 
by gauging the symmetry $U(1,D-1)$ instead of the usual $SO(1,D-1)$ 
 \cite{cham}. The  metric contains  the usual symmetric part 
plus an  antisymmetric tensor whose origin traces back to the
fact that the non-commutativity antisymmetric
 parameter $\Theta_{\mu\nu}$ becomes
a dynamical field. 
This might not be necessarily a threat to
our construction  since non-commutativity shows  up only in the two
extra dimensions. Furthermore, one can write a  both unique and gauge invariant
gravity action \cite{cham}.
Clearly, all these issues  deserve to be addressed and we are now working along
these directions. What makes us hopeful is that the study of the
gravitational background and its excitations in the presence of the
non-commutative soliton is to some extent simplified by the 
nice properties of the non-commutative soliton and by its
capability of playing the role of a warp factor.

It would be also worth considering 
what are  the cosmological implications of our set-up. 
Because of the UV/IR mixing, 
fluctuations of a scalar field at small scales $\ell$ 
reappear at large distances $\Theta/\ell$ \cite{con} and it has been
recently shown  that
 solitons can travel faster than the speed of 
light for arbitrarily long distances \cite{light}. These ingredients may be 
relevant  to construct an alternative to the
theory of inflation \cite{review} and  the generation
of the cosmological density perturbations  \cite{chu}. Another 
interesting project might be 
to investigate if there are 
new types of four-dimensional topological
defects whose  core might act as
windows to the extra dimensions \cite{topological} and thus to
the non-commutative world.

\vskip2cm
\centerline{\large\bf Acknowledgments}
\vskip 0.2cm

We would like to thank J.F.L. Barbon and T. Gherghetta for useful discussions.

\end{document}